\documentclass[final,5p,times,twocolumn]{elsarticle}

\usepackage{amssymb}

\usepackage{dcolumn}
\usepackage{bm}
\biboptions{sort&compress}

\journal{Nuclear Instruments and Methods in Physics Research Section A}

\begin{document}

\begin{frontmatter}



\title{Multi-color, femtosecond $\gamma$-ray pulse trains driven by comb-like electron beams}


\author[1]{S. Y. Kalmykov\corref{cor1}}\ead{s.kalmykov.2013@ieee.org}\author[2]{X. Davoine}\author[3]{I. Ghebregziabher}\author[1]{B. A. Shadwick}
\cortext[cor1]{Corresponding author.\ All correspondence must be addressed to Serge Y.\ Kalmykov, Leidos Inc., 2109 Air Park Rd.\ SE, Ste.\ 200, Albuquerque, NM 87106, USA.\ Tel.: +1 505 433 4948.}
\address[1]{Department of Physics and Astronomy, University of Nebraska -- Lincoln, Lincoln, NE 68588-0299, USA}
\address[2]{CEA, DAM, DIF, Arpajon F-91297, France}
\address[3]{Penn State Hazleton, Hazleton, PA 18202, USA}

%

\begin{abstract}
Photon engineering can be exploited to control the nonlinear evolution of the drive pulse in a laser-plasma accelerator (LPA), offering new avenues to tailor electron beam phase space on a femtosecond time scale.\ One promising option is to drive an LPA with an incoherent stack of two sub-Joule, multi-TW pulses of different colors. Slow self-compression of the bi-color optical driver delays electron dephasing, boosting electron beam energy without accumulation of a massive low-energy tail.\ The modest energy of the stack affords kHz-scale repetition rate at manageable laser average power.\ Propagating the stack in a pre-formed plasma channel induces periodic self-focusing in the trailing pulse, causing oscillations in the size of accelerating bucket.\ The resulting periodic injection generates, over a mm-scale distance, a train of GeV-scale electron bunches with 5D brightness exceeding $10^{17}$ A/m$^2$.\ This unconventional comb-like beam, with femtosecond synchronization and controllable energy spacing of components, emits, via Thomson scattering, a train of highly collimated gigawatt $\gamma$-ray pulses.\ Each pulse, corresponding to a distinct energy band between 2.5 and 25 MeV, contains over $10^6$ photons.
\end{abstract}

\begin{keyword}


Laser plasma acceleration \sep Thomson scattering  \sep laser pulse stacking \sep blowout \sep electron trapping \sep comb-like beam

\PACS 41.75.Jv \sep 41.85.Ct \sep 42.65.Jx \sep 42.65.Wi \sep 52.35.Mw \sep 52.38.Hb \sep 52.38.Kd \sep 52.38.Ph
\end{keyword}

\end{frontmatter}

\newpage
\section{Introduction}
\label{Intro}
A train of electron bunches of different energies, termed  the ``comb-like beam'' \cite{Pertillo_PRAB_2014,Cianchi_PRAB_2015,Shevelev_PRAB_2017}, is a promising source of multi-color X-ray radiation \cite{Pertillo_PRAB_2014}, with applications in diagnostics of ultra-fast processes \cite{Cahoon_Sci_2008,Scoby_PhD_2008} and radiation medicine \cite{Willekens_ESR_2011}.\ Multi-kA comb-like beams may also effectively excite plasma wakefields \cite{Muggli_PRL_2008,Pompili_NIMA_2016}.\ Free-electron lasers \cite{Anania_NJP_2014,Petralia_PRL_2015} and THz sources \cite{Shen_PRL_2011} may take advantage of tailored electron beams (e-beams), modulated both in time and energy, to customize emission bandwidth and temporal properties.

Conventional linear accelerators generate trains of 100 fs bunches synchronized on a picosecond time scale \cite{Cianchi_PRAB_2015,Shevelev_PRAB_2017,Muggli_PRL_2008,Pompili_NIMA_2016,Anania_NJP_2014}.\ Laser-plasma accelerators (LPAs) \cite{Malka_PoP_2012} access much shorter scales, generating kA-scale, sub-10 fs single electron bunches \cite{Lundh_Nat_Phys_2011}, or trains of such bunches \cite{Lundh_PRL_2013,Walker_NJP_2013,Manahan_NJP_2014}.\ These trains, formed by particles accelerated in consecutive wake buckets \cite{Lundh_PRL_2013}, or by multiple injection events into the first bucket \cite{Walker_NJP_2013,Manahan_NJP_2014}, are accessible to all-optical control \cite{Kalmykov_PoP_2015,Kalmykov_PPhCF_2016,Kalmykov_AIP_Proc_2017}.\ This control favors generation of X-rays via Thomson scattering \cite{Pertillo_PRAB_2014,Esarey_PRE_1993,Ride_PRE_1995,Tomassini_PRAB_2013,Chen_PRL_2013,Ghebregziabher_PRAB_2013,Powers_NatPhoton_2014,Liu_OptLett_2014}.

Thomson scattering (TS), the low-energy semi-classical limit of the general quantum-mechanical inverse Compton scattering process, occurs in a collision of a relativistic electron with an interaction laser pulse (ILP). An electron with a Lorentz factor $\gamma_e \gg 1$, propagating at an angle to the ILP, experiences the compressed wave front of the pulse, the maximum compression occurring in the direction of the particle propagation. While oscillating in the ILP field, the electron emits radiation, scattering the compressed wave front. An observer in the far field detects an angular distribution of high-energy photons, with the energy being the highest for a detector placed in the direction of electron propagation. For the head-on collision, the ILP photon energy is Doppler up-shifted by a factor $4\gamma_e^2$. A 900 MeV electron thus converts 1.5 eV ILP photons into 19 MeV $\gamma$-photons.

A realistic e-beam, with nonzero divergence and energy spread, imprints its six-dimensional (6D) phase space onto the TS $\gamma$-ray emission pattern and energy spectrum. TS may thus serve as a diagnostic of electron bunching in momentum space \cite{Kalmykov_PoP_2015,Kalmykov_PPhCF_2016,Kalmykov_AIP_Proc_2017,Hsu_PRE_1996}. This bunching is highly susceptible to the details of the drive laser pulse evolution in the course of acceleration. The latter may be deliberately controlled through modification of the drive pulse phase, thus affording considerable freedom in e-beam shaping \cite{Kalmykov_PoP_2015,Mangles_APL_2009,Popp_PRL_2010,Kalmykov_NJP_2012}.

This report proposes a way of producing trains of quasi-monochromatic, high-flux, femtosecond $\gamma$-ray pulses via TS from GeV-scale, tunable comb-like e-beams.\ The element of innovation is in producing these e-beams in an LPA driven with a Joule-scale stack of pulses with large difference frequency.\ Modern optical engineering permits obtaining the necessary large shift in wavelength ($\Delta \lambda \! \sim \! \lambda_0 \! \sim 1$ $\mu$m) using a Raman cell, with subsequent conventional chirped-pulse amplification \cite{Grigsby_JOSAB_2008,Vicario_OL_2016}, or via energy-efficient methods of frequency-doubling.\ The modest peak power/energy of the stacked driver affords kHz-scale repetition rate with kW average power; a hard yet practical task \cite{Gizzi_ApplSci_2013}.\ High repetition rate is critical for applications dependent on dosage.\ It also enables real time optimization of the laser-plasma interaction using adaptive optics and genetic algorithms \cite{He_NatComm_2015}.\ Further, the significant reduction in the size of the plasma (in comparison to the predictions of accepted scaling \cite{Lu_PRAB_2007}) facilitates predictive high-fidelity three-dimensional particle-in-cell (PIC) simulations \cite{Cowan_JPP_2012}.

Advancing in time the blue-shifted stack component (i.e.\ introducing a piecewise negative frequency chirp) compensates for the non-linear red-shift imparted by the plasma wake at the stack leading edge. This slows down self-compression of the optical driver, mitigating the primary effect that limits electron energy gain \cite{Lu_PRAB_2007}. By delaying electron dephasing and taking advantage of the TV/m-scale accelerating gradient of a dense plasma ($n_0 \sim 10^{19}$ cm$^{-3}$), one may expect a GeV gain over a mm-scale distance, without resorting to PW-scale peak power lasers \cite{Kalmykov_PoP_2015}. In addition, as the stack compresses slowly, massive continuous injection is avoided. The electron spectra remain free of a high-charge, low-energy background \cite{Kalmykov_PoP_2015,Kalmykov_NJP_2012}.

As the rigid head of the stack, resilient to red-shift and self-compression, drives the wake, the tail, residing in a soft plasma channel -- the bubble of electron density -- defines the dynamics of electron injection.\ Unbalanced self-focusing in the tail induces oscillations in the bubble size, causing periodic injection, modulating the e-beam current on a femtosecond scale \cite{Kalmykov_PPhCF_2016}.\ This effect may be achieved by weakly focusing the tail \cite{Kalmykov_AIP_Proc_2017} or by propagating the stack in a pre-formed plasma channel \cite{Kalmykov_PoP_2015}.\ In this report, we show that the comb-like e-beam, generated in the regime of Ref.\ \cite{Kalmykov_PoP_2015}, is an ideal source of a multi-pulse TS $\gamma$-ray beam, featuring up to four distinct energy bands in the range of interest to nuclear photonics, 2.5 -- 25 MeV \cite{Rykovanov_JPB_2014}, with up to $10^6$ photons in each band.

Section \ref{Sec1} of this paper illustrates physics of periodic injection in the channel and quantifies the resulting GeV comb-like e-beam.\ Section \ref{Sec2} demonstrates control over the spectral content and bandwidth of the TS $\gamma$-ray pulse trains, emitted from the comb-like e-beam.\ Section \ref{Conclusion} summarizes the results.

\section{Stack-driven LPA in a leaky plasma channel}
\label{Sec1}

The reported study is accomplished via quasi-cylindrical PIC simulations using the code CALDER-Circ \cite{Lifschitz_JCP_2009}.\ Its numerical Cherenkov-free electromagnetic solver \cite{Lehe_PRAB_2013} maintains negligibly low numerical dispersion, avoiding numerical emittance dilution.\ The numerical aspects are those of Ref.\ \cite{Kalmykov_PoP_2015}.

\begin{figure}[t]
\centering
\includegraphics[scale=1]{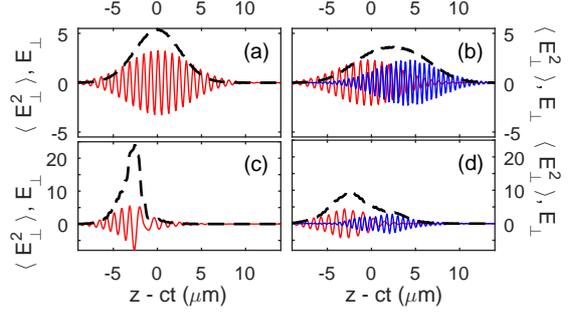}
\caption{Frequency shift between stack components makes the stack resilient to self-compression.\ Panels (a), (c): the reference case.\ Panels (b), (d): the stack with the delay $T \! = \! 15$ fs and frequency ratio $\Omega \! = \! 1.5$.\ The pulses propagate to the right; $z \! = \! ct$ is the centroid of the carrier-frequency component ($E_0$) in vacuum.\ The top row shows the fields at the plasma entrance ($z \! = \! 0$).\ Panels (c) and (d) correspond to $z \! = \! 1.6$ mm (the point of full compression in the reference case) and $z = 2.15$ mm, respectively.\ Snapshots of electric field, in units of $(m_e \omega_0 c)/|e| \! = \! 4$ TV/m, are taken on-axis; in the simulation, $\mathbf{E}_0 \perp \mathbf{E}_\mathrm{head}$.\ Red (dark gray): $E_0$.\ Blue (gray in panel (b)): $E_\mathrm{head}$.\ Dashed curve: $\langle E_\perp^2 \rangle \! = \! \langle E_0^2 \rangle \! + \! \langle E_\mathrm{head}^2 \rangle$, where $\langle \cdots \rangle$ denotes averaging over an optical cycle.}\label{fig:figure1}
\end{figure}

We fix the total laser energy at 1.4 J. This energy may be concentrated in a single, transform-limited, linearly polarized Gaussian pulse with a carrier wavelength $\lambda_0 \! = \! 0.805$ $\mu$m and full width at half-maximum in intensity $\tau_L \! = \! 20$ fs, which defines the reference case \cite{Kalmykov_PoP_2015}.\ The plasma begins at $z \! = \! 0$ with a 0.5 mm linear ramp, followed by a longitudinally uniform section with the electron density $n_0 \! = \! 6.5 \! \times \! 10^{18}$ cm$^{-3}$.\ The electric field of the pulse, focused at the plasma border into a spot $r_0 \! = \! 13.6$ $\mu$m, is given by $\mathbf{E}_0 (x,y,z \! = \! 0,t) \! = \! \mathbf{e}_x (m_e\omega_0 c /|e|)\,\mathcal{E}_0 \exp(-\mathrm{i}\omega_0t \!  - \! 2\ln 2\, t^2/\tau_L^2 \! - \! r^2/r_0^2)$.\ Here, $r^2 \! = \! x^2 \! + \! y^2$, $\mathbf{e}_x$ is the unit polarization vector, $\omega_0 \! = \! 2\pi c /\lambda_0$, $c$ is the vacuum speed of light, and $-|e|$ and $m_e$ are electron charge and rest mass, respectively.\ With these parameters, the normalization factor is $m_e\omega_0 c /|e| \! = \! 4$ TV/m.\ The single 70 TW pulse ($\mathcal{E}_0 \! = \! 3.27$) depletes as soon as the electrons reach dephasing, promising to maximize acceleration efficiency \cite{Lu_PRAB_2007}.\ However, this strategy leads to a copious dark current and overall low beam quality \cite{Kalmykov_PoP_2015}.\ Figure \ref{fig:figure1}(c) shows that the reference pulse self-compresses into an optical shock with a sub-cycle-length rising edge. Electron density pileup inside the shock causes continuous expansion of the bubble, which is a purely quasistatic process, almost unaffected by the beam loading \cite{Kalmykov_NJP_2012}.\ The expanding bubble sucks in electrons from its sheath, building up a massive energy tail containing up to 80\% of energetic particles \cite{Kalmykov_PoP_2015,Kalmykov_NJP_2012}.

\begin{figure}[t]
\centering
\includegraphics[scale=1]{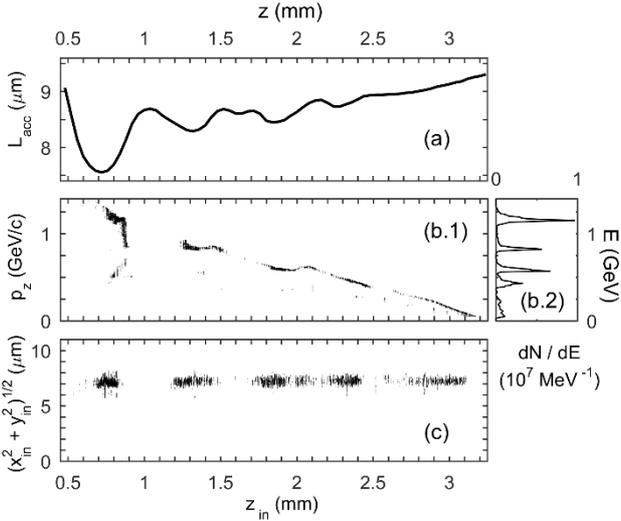}
\caption{Periodic injection in a plasma channel yields a four-component e-beam, with almost no tail. Longitudinal momenta vs initial longitudinal positions (b.1) and initial positions (c) are presented of the electrons with $E \! > \! 50$ MeV, crossing the plane $z = 3.24$ mm. Oscillations in the bubble size, shown in (a), produce four consecutive injections, forming the energy comb (b.2).} \label{fig:figure2}
\end{figure}

Blue-shifting the leading edge of the optical driver compensates for the nonlinear red-shift imparted by the wake, making the driver resilient to self-compression.\ A practical approach is to split the pulse energy evenly between two pulses of the same duration, increase the frequency of one of them, and advance the up-shifted pulse in time by a fraction of the pulse length \cite{Kalmykov_PoP_2015}.\ The electric field of this incoherent stack is $\mathbf{E}_\mathrm{stack} (z \! = \! 0) \! = \! 2^{-1/2}\mathbf{E}_0 \! + \! \mathbf{E}_\mathrm{head}$, where $\mathbf{E}_\mathrm{head} \! = \! \mathbf{e}_y (m_e \omega_0 c/|e|)\, \mathcal{E}_\mathrm{head} \exp [-\mathrm{i} \omega_\mathrm{head} (t \! + \! T) \! - \! 2 \ln 2 (t \! + \! T)^2/\tau_L^2 \! - \! r^2/r_0^2]$.\ Here, $\mathcal{E}_\mathrm{head} \! = \! \mathcal{E}_0/\sqrt{2}  \! = \! 2.31$, $\mathbf{e}_y $ is the unit polarization vector, the delay $T \! > \! 0$, and $\Omega \! = \! \omega_\mathrm{head}/\omega_0 \! > \! 1$.\ Figures \ref{fig:figure1}(b) and \ref{fig:figure1}(d) correspond to the optimal case of $T \! = \! 15$ fs and $\Omega \! = \! 1.5$ ($\lambda_\mathrm{head} \! = \! 2\lambda_0/3 \! \approx \! 0.535$ $\mu$m).\ In stark contrast to the reference scenario, compression of the stack is minimal, which leads to doubling electron energy compared to predictions based on scalings (885 vs 420 MeV) and the near-elimination of the low-energy tail \cite{Kalmykov_PoP_2015}.\ Varying the delay and frequency ratio in the stack permits considerable flexibility in tuning the parameters of quasi-monoenergetic (QME), GeV-scale e-beams.

\begin{figure}[t]
\centering
\includegraphics[scale=0.93]{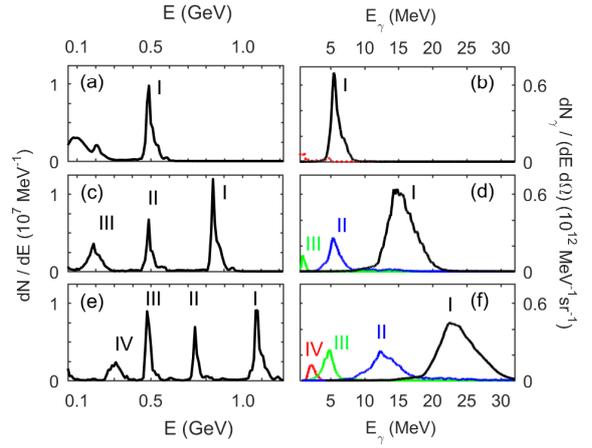}
\caption{Photon energy comb emitted by a train of e-bunches. Left column: electron energy spectra. Right column:  TS $\gamma$-ray flux, split into partial signals from individual bunches (on axis, in the direction of e-beam propagation). Spectra of individual e-bunches I -- IV (and corresponding $\gamma$-ray beamlets) are marked in the order of their decreasing energy. The e-beams are extracted from the CALDER-Circ simulation to compute the TS spectra when the highest-energy bunch reaches (a), (b) $z = 1.5136$ mm; (c), (d) $z = 2.2308$ mm; (e), (f) $z = 2.9082$ mm. Tables \ref{tab:a} and \ref{tab:b} contain statistics of individual bunches and corresponding TS $\gamma$-ray signals.} \label{fig:figure3}
\end{figure}

Propagating the stack in a leaky channel \cite{Antonsen_PRL_1995,Milchberg_PoP_1996}
\begin{displaymath}
n_e(r) = \left\{ \begin{array}{ll}
n_0(1+r^2/r_\mathrm{ch}^2) & \textrm{for $r \le  r_\mathrm{ch}$,}\\
2n_0(2-r/r_\mathrm{ch})    & \textrm{for $r_\mathrm{ch} < r \leq 2 r_\mathrm{ch}$,}\\
0 & \textrm{for $r > 2 r_\mathrm{ch}$,}
\end{array} \right.
\end{displaymath}
with $r_\mathrm{ch} \! \approx \! 1.25 r_\mathrm{m}$, where $r_\mathrm{m} \! = \! \omega_\mathrm{pe}^2 r_0^2 /(2c) \! \approx \! 44$ $\mu$m is the radius of the channel matched for single-mode guiding\footnote{In the linear regime, the channel supports only the lowest mode of propagation while allowing higher order modes to leak away.}, and $\omega_\mathrm{pe} \! = \! \sqrt{4\pi e^2 n_0/m_e}$ is the plasma frequency on axis, modulates the e-beam current, forming the energy comb seen in Fig.\ \ref{fig:figure2}(b.2).

Figure \ref{fig:figure2} relates the details of bubble evolution to the kinetics of periodic injection.\ The bubble size, defined as the length of the accelerating phase on axis (i.e.\ the region inside the bubble where the longitudinal electric field is negative), is shown in Fig.\ \ref{fig:figure2}(a) versus propagation distance.\ Figure \ref{fig:figure2}(b.1) presents the collection phase space (longitudinal momentum versus initial longitudinal position) of electrons with $E \! > \! 50$ MeV, crossing the plane $z \! = \! 3.24$ mm, correlating injection intervals with the intervals of bubble expansion.\ Initial positions of energetic electrons exiting the plasma, shown in Fig.\ \ref{fig:figure2}(c), reveal the same correlation while confirming the absence of on-axis injection.

The channel upsets the balance between the radiation pressure in the stack tail and the radial restoring force due to charge separation in the bubble.\ The tail flaps, driving the bubble boundaries sidewards, inducing oscillations\footnote{Experimentally detectable with a frequency-domain streak camera \cite{Downer_PRL_2014}.} in the bubble size synchronized with the oscillations in the tail spot size \cite{Kalmykov_PPhCF_2016}.\ Every expansion of the bubble injects electrons from the sheath.\ Subsequent contraction clips the bunch, limiting the charge to tens of pC while preserving a 100 kA-scale current, and monochromatizes the bunch through rapid phase space rotation \cite{Kalmykov_PoP_2011}, leading to the formation of four narrow energy bands containing 10\% of the stack energy.\ The lowest-energy band in Fig.\ \ref{fig:figure2}(b.2), with $\langle E \rangle \! \approx \! 420$ MeV, corresponds to the maximum possible energy gain in the reference case, while the highest-energy one exceeds this gain by a factor 2.75.\ As the electrons approach dephasing ($z \! > \! 2.7$ mm), the bubble expands very slowly, keeping the energy tail ($E \! < \! 250$ MeV) minimal.

\begin{table}[b]
\centering
	\begin{tabular}{l r r r r }
\hline
Quantity   & Q (pC) & $\sigma_\tau$ (fs) & $\langle I \rangle$ (kA) &  $\Delta T_n$ (fs) \\[0.5ex]

\hline
Beam I          & 63.8  &  0.63         & 101                & 2.15 \\
Beam II         & 38.1  &  0.74         & 51.7               & 3.15 \\
Beam III        & 35.6  &  1.10         & 32.5               & 3.67 \\
Beam IV         & 26.2  &  0.64         & 41.0               & \\
\hline
\end{tabular}
\caption{Parameters of QME bunches making up the energy combs in Figs.\ \ref{fig:figure3}(a), \ref{fig:figure3}(c), \ref{fig:figure3}(e). $Q$ is the charge; $\sigma_\tau$ is the RMS length; $\langle I \rangle = Q/\sigma_\tau$ is the mean current; $\Delta T_n = (\langle z_n \rangle - \langle z_{n+1} \rangle)/c$ is the delay between adjacent bunches.}
\label{tab:a0}
\end{table}

The energy spectra shown in Figs.\ \ref{fig:figure3}(a), \ref{fig:figure3}(c), and \ref{fig:figure3}(e) demonstrate that limiting acceleration length controls the number of the bunches in the train, their mean energy, and energy spacing.\ Table \ref{tab:a0} shows that these sub-fs bunches are synchronised on a few-fs time scale.\ Separating them in a magnetic spectrometer and using delay lines may adjust synchronization time to the demands of specific applications. The phase space of individual bunches is characterized by the root-mean-square (RMS) divergence $\sigma_\alpha$ and normalized transverse emittance, $\varepsilon_\perp^N$ \cite{Kalmykov_PoP_2015}. The latter, in combination with the mean current, $\langle I \rangle$, defines the five-dimensional (5D) brightness, $B_n = 2\langle I \rangle(\pi\varepsilon_\perp^{\, N})^{-2}$ \cite{Cianchi_NIMA_2016}. Table \ref{tab:a} shows evolution of these quantities in the course of acceleration. Thanks to the numerical Cherenkov-free electromagnetic solver in CALDER-Circ \cite{Lehe_PRAB_2013}, the sub-$\mu$m emittance of the bunches varies by less than 5\%. In combination with 30 -- 100 kA current, this gives $B_n > 10^{17}$ A/m$^2$, promising high efficiency in emission of high-flux TS $\gamma$-ray pulses \cite{Cianchi_NIMA_2016}.

\begin{table}[b]
\centering
	\begin{tabular}{l r r r r r r}
\hline
Quantity     & $\langle E \rangle$  & $\sigma_E$  & $\sigma_\alpha$ & $\varepsilon^{\, N}_\perp$  & $B_n$              &  W \\[0.5ex]
Unit         & MeV                  & MeV         & mrad            & $\mu$m                 &   A/m$^2$     & mJ \\[0.5ex]
\hline
\hline
I\,\,\,\, a      & 509                & 28.5        &  1.70          & 0.35                  & $1.65 \! \times \! 10^{17}$  & 32.5 \\
I\,\,\,\, c       & 863                & 27.2        &  1.30          & 0.36                   & $1.60\! \times \! 10^{17}$  & 55.0 \\
I\,\,\,\, e       & 1095              & 31.0        &  0.97          & 0.38                   & $1.39\! \times \! 10^{17}$  & 70.0 \\
\hline
II\,\, c          & 546                & 93.9        &  1.21          & 0.27                   & $1.48\! \times \! 10^{17}$  & 20.8 \\
II\,\, e           & 791                & 91.0        &  1.25          & 0.28                   & $1.30\! \times \! 10^{17}$  & 30.2 \\
\hline
III c               & 199                & 38.0        &  3.35          & 0.25                   & $1.05\! \times \! 10^{17}$  & 7.08 \\
III e                & 498                & 18.6        &  1.71          & 0.29                   & $0.76\! \times \! 10^{17}$  & 17.7 \\
\hline
IV e                 & 316                & 30.9        &  2.50           & 0.23                   & $1.54\! \times \! 10^{17}$  & 8.28 \\
\hline
\end{tabular}
\caption{Evolution of mean energy $\langle E \rangle$; energy variance $\sigma_E$; RMS angular spread, $\sigma_\alpha$; normalized transverse emittance, $\varepsilon^N_\perp$; 5D brightness, $B_n$; and total energy $W$ of the QME bunches making up the energy combs shown in Figs.\ \ref{fig:figure3}(a), \ref{fig:figure3}(c), and \ref{fig:figure3}(e). }
\label{tab:a}
\end{table}

\section{Polychromatic $\gamma$-rays from Thomson scattering}
\label{Sec2}

The geometry of e-beam collision with a linearly polarized, paraxial interaction laser pulse (ILP) and the method of computation of Thomson scattering spectra from individual bunches \cite{Ghebregziabher_PRAB_2013} are identical to those of Refs.\ \cite{Kalmykov_PPhCF_2016,Kalmykov_AIP_Proc_2017}.\ The ILP has a 0.8 $\mu$m carrier wavelength (photon energy $E_\mathrm{int} \! = \! 1.5$ eV), 250 fs duration (0.3\% FWHM bandwidth in spectral intensity), and 16.8 $\mu$m waist size (Rayleigh length 1.1 mm).\ Its modest normalized vector potential, $a_{\mathrm{int}} \! = \!  0.1$, ensures linearity of interaction~\cite{Tomassini_PRAB_2013}.

\begin{table}[b]
\centering
	\begin{tabular}{l r r r r r}
\hline
Quantity            & $\langle E_\gamma \rangle$  & $\Delta E_\gamma$  & $\Omega_\gamma$   & $N_\Omega$ &  $W_\Omega$ \\[0.5ex]
Unit                & MeV  & MeV  & $\mu$sr  & $10^6$ & $\mu$J \\[0.5ex]
\hline
\hline
I\,\,\,\,\, b               & 5.84   & 0.99  & 1.58  & 1.60    &  1.49   \\
I\,\,\,\,\, d               & 15.5   & 2.31  & 0.55  & 1.64    &  4.05   \\
I\,\,\,\,\, f               & 24.1   & 3.81  & 0.34  & 1.62    &  7.42   \\
\hline
II\,\,\, d              & 5.57   & 1.07  & 1.38  & 0.61    &  0.55   \\
II\,\,\, f              & 12.8   & 2.00  & 0.65  & 0.65    &  1.33   \\
\hline
III\, d             & 0.96   & 0.30  & 10.3  & 0.82    &  0.13   \\
III\, f             & 4.67   & 0.93  & 1.65  & 0.58    &  0.44   \\
\hline
IV\, f              & 2.27   & 0.51  & 4.12  & 0.60    &  0.22   \\
\hline
\end{tabular}
\caption{Statistics of quasi-monochromatic $\gamma$-ray beamlets from Fig.\ \ref{fig:figure3}, panels (b), (d), and (f). $\langle E_\gamma \rangle$ is the mean energy; $\Delta E_\gamma$   is the energy variance; $N_\Omega$ and $W_\Omega = N_\Omega\langle E_\gamma \rangle$ are the number of photons and energy radiated into the observation solid angle $\Omega_\gamma = (\pi/2)\langle\gamma_e\rangle^{-2}$ in the direction of e-beam propagation.}
\label{tab:b}
\end{table}

The photon flux (on axis, in the direction of e-beam propagation), corresponding to the QME features in Figs.\ \ref{fig:figure3}(a), \ref{fig:figure3}(c), and \ref{fig:figure3}(e), is shown in Figs.\ \ref{fig:figure3}(b), \ref{fig:figure3}(d), and \ref{fig:figure3}(f), respectively.\ The $\gamma$-ray spectra contain up to four quasi-monochromatic bands, centered about $\langle E_\gamma\rangle \! \approx \! 4\langle\gamma_e\rangle E_\mathrm{int}$, with an energy spread of 15 to 20\%, as indicated by the entries in Table \ref{tab:b}.\ TS simulations with a reduced phase space of individual bunches \cite{Kalmykov_AIP_Proc_2017} prove that this bandwidth is imparted primarily by the energy spread in the bunches ($\approx \! 30$ and 90 MeV for bunches I and II, almost constant throughout acceleration).\ Focusing with highly chromatic magnetic quadrupole lenses \cite{Weingartner_PRAB_2011} may select electrons from much narrower energy intervals before their collision with the ILP, allowing for the possibility of sub-percent $\gamma$-ray pulse bandwidth, while reducing the number of photons in the band.\ Adding a frequency chirp to the ILP may be another practical way to reduce the $\gamma$-ray pulse bandwidth \cite{Ghebregziabher_PRAB_2013}.\ Given the unconventional shape of electron momentum chirp in LPA-produced beams, such as seen, e.g.\ in Figures 3(a) of Ref.\ \cite{Kalmykov_AIP_Proc_2017}, this topic deserves special consideration, and will be discussed elsewhere.

The collimation of high-energy $\gamma$-photons, as well as the number of photons in the observation cone corresponding to an individual band, are important metrics for applications.\ It should be noted (cf.\ the entries in Table \ref{tab:a}) that the divergence of electron bunches, $\sigma_\alpha$, exceeds $\langle\gamma_e\rangle^{-1}$ by a factor 1.5 to 2.\ To evaluate the reduction in photon energy and flux with an increase in the observation angle (viz.\ to estimate the effective apex angle of the emission cone), we simulate TS from an individual bunch, gradually increasing the observation angle, from $\theta \! = \! 0$, corresponding to the detector placed on the bunch propagation axis, to $\theta \! = \! \sigma_\alpha$.\ As $\theta$ increases from 0 to $\langle \gamma_e\rangle^{-1}$, the mean photon energy drops by 25\%, while the photon flux drops ten-fold.\ To a good approximation, all photons with energies above $3\langle E_\gamma\rangle/4$ are emitted into the observation cone of apex angle $2\theta \! = \! 2\langle\gamma_e\rangle^{-1}$.\ To estimate the number of QME high-energy photons scattered in the direction of bunch propagation, we conservatively choose the observation solid angle of the band $\Omega_\gamma \! = \! (\pi/2)\langle\gamma_e\rangle^{-2}$, i.e.\ the solid angle of the cone with an apex angle $2\theta \! = \! \sqrt{2}\langle\gamma_e\rangle^{-1}$.\ Then we take the photon flux corresponding to backscattering ($\theta \! = \! 0$), integrate it over the energy, and multiply the result by $\Omega_\gamma$.\ Corresponding entries $N_\Omega$ in Table \ref{tab:b} are of the order $10^6$ photons per band.\ As a consequence of electron bunch emittance preservation, the number of photons in bands I and II remains almost constant as their mean energy increases, with the power of the collimated, quasi-monochromatic, femtosecond photon pulses reaching 12 and 1.75 GW, respectively.

Given the modest energy of the stacked driver, a kW average power amplifier \cite{Gizzi_ApplSci_2013} would enable an LPA with a kHz-scale repetition rate. Increasing the ILP length by an order of magnitude (to 2.5 ps) would further increase the photon gain without compromising $\gamma$-ray bandwidth. The expected gain of $10^{10}$ ph/s is sufficiently high to make a tunable, multi-color TS source attractive for a non-destructive inspection system for special nuclear materials. From simulations of Ref.\  \cite{Ohgaki_TNS_2017}, based on the data of their detection experiment, this flux may be sufficient to identify a nuclear resonance fluorescence peak from a 1 kg of highly enriched uranium within a few minutes.

\section{Summary and outlook}
\label{Conclusion}

In an LPA, electrons self-injected from the ambient plasma are accelerated in a cavity of electron density maintained by the radiation pressure of a relativistically intense laser pulse.\ Deformations in this cavity, caused by the deformations of the optical driver, can be controlled through photon engineering, in order to impart desirable modulations in the e-beam current.\ An incoherent stack of two pulses, the blue-shifted one advanced in time by a fraction of its length, is resilient to nonlinear self-compression \cite{Kalmykov_NJP_2012,Kalmykov_PoP_2015}.\ This avoids contamination of the electron spectrum with a high-charge, low-energy tail, also increasing the energy of the dominant QME component far beyond predictions of the accepted scaling.\ The Joule-scale energy requirements of the stack affords kHz-scale repetition rate at a manageable average power, favoring applications dependent on dosage.\  The presented case study demonstrates that propagating the stack in a leaky plasma channel induces oscillations in the size of accelerating cavity, creating a background-free comb-like e-beam, with the highest-energy bunch reaching 1.2 GeV (compared to 420 MeV predicted by scaling), all bunches having 5D brightness above $10^{17}$ A/m$^2$.

Thomson scattering from comb-like e-beam produces a train of synchronized X-ray pulses corresponding to distinct energy bands \cite{Pertillo_PRAB_2014,Kalmykov_PPhCF_2016,Kalmykov_AIP_Proc_2017}, to the benefit of radiation medicine \cite{Willekens_ESR_2011} and national security \cite{Rykovanov_JPB_2014,Ohgaki_TNS_2017}.\ In our case, TS generates $\gamma$-rays in up to four distinct bands in the range 2.5 MeV $<\langle E_\gamma\rangle<$ 25 MeV.\ The number of bands, their mean energy, and the energy spacing may be controlled by changing the plasma length, using conventional LPA target design \cite{Vargas_APL_2014}.\ This tunability is an asset to applications in nuclear photonics \cite{Rykovanov_JPB_2014,Ohgaki_TNS_2017}.\ Each band, having 15 to 20\% energy spread, imparted by a few-percent energy spread in the electron bunches, corresponds to femtosecond-length, gigawatt $\gamma$-radiation pulse containing up to $10^6$ photons in the micro-steradian observation solid angle.\ Femtosecond length and synchronization of $\gamma$-ray pulses and electron bunches are useful for nuclear pump-probe experiments \cite{Scoby_PhD_2008}.


\section*{Acknowledgements}
The work of SYK and BAS was supported in part by the National Science Foundation Grant PHY-1535678.  Thomson scattering simulations were completed utilizing the Holland Computing Center of the University of Nebraska.

\end{document}